\documentclass[aps,prb,amsmath,twocolumn,groupedaddress,superscriptaddress,citeautoscript,showpacs,floatfix]{revtex4-1}
\usepackage[dvips]{graphicx}
\usepackage{multirow}
\begin{document}

\title{Ideal strength of random alloys from first-principles theory}
\author{Xiaoqing Li}
\email{xiaoqli@kth.se}
\affiliation{Applied Materials Physics, Department of Materials Science and Engineering, Royal Institute of Technology, Stockholm SE-10044, Sweden}
\author{Stephan Sch\"onecker}
\email{stesch@kth.se}
\affiliation{Applied Materials Physics, Department of Materials Science and Engineering, Royal Institute of Technology, Stockholm SE-10044, Sweden}
\author{Jijun Zhao}
\email{zhaojj@dlut.edu.cn}
\affiliation{School of Physics and Optoelectronic Technology and College of Advanced Science and Technology, Dalian University of Technology, Dalian 116024, China}
\affiliation{Key Laboratory of Materials Modification by Laser, Electron, and Ion Beams (Dalian University of Technology), Ministry of Education, Dalian 116024, People's Republic of China}
\author{B\"{o}rje Johansson}
\affiliation{Applied Materials Physics, Department of Materials Science and Engineering, Royal Institute of Technology, Stockholm SE-10044, Sweden}
\affiliation{Department of Physics and Astronomy, Division of Materials Theory, Uppsala University, Box 516, SE-75120, Uppsala, Sweden}
\author{Levente Vitos}
\affiliation{Applied Materials Physics, Department of Materials Science and Engineering, Royal Institute of Technology, Stockholm SE-10044, Sweden}
\affiliation{Department of Physics and Astronomy, Division of Materials Theory, Uppsala University, Box 516, SE-75120, Uppsala, Sweden}
\affiliation{Research Institute for Solid State Physics and Optics, Wigner Research Center for Physics, Budapest H-1525, P.O. Box 49, Hungary}

\date{\today}

\begin{abstract}
The all-electron exact muffin-tin orbitals method in combination with the coherent-potential approximation was employed to investigate the ideal tensile strengths of elemental V and Mo solids, and V- and Mo-based random solid solutions.
Under uniaxial $[001]$ tensile loading, the ideal tensile strength of V is $11.6$ GPa and the lattice fails by shear. Assuming isotropic Poisson contraction, the ideal tensile strength are $26.7$ and $37.6$ GPa for V in the $[111]$ and $[110]$ directions, respectively.  The ideal strength of Mo is $26.7$ GPa in the $[001]$ direction and decreases when a few percent Tc are introduced in Mo. For the V-based alloys, Cr increases and Ti decreases the ideal tensile strength in all principal directions. Adding the same concentration of Cr and Ti to V leads to ternary alloys with similar ideal strength values as that of pure V. The alloying effects on the ideal strength are explained using the electronic band structure.
\end{abstract}


\maketitle

\section{\label{sec:introduction}Introduction}

High strength and good ductility define the two most important mechanical properties of metallic structure materials.~\cite{property:1} In materials, strength is usually controlled by the occurrence of grain boundaries, cracks, dislocations and other micro-structural defects. If such defects were not present, the strength would be limited by the stress at which the lattice itself becomes unstable with respect to a homogeneous  strain. This stress referred to as the ideal strength, is the possible maximum strength of an ideal single crystal. The ideal strength is an inherent property of a material which can offer insight into the correlation between the intrinsic chemical bonding and the crystal symmetry and has been accepted as an essential mechanical parameter of single crystal materials.~\cite{inherent:property}

The experimental data on ideal strength are rather limited, since it is difficult to measure the ideal strength using experimental tools. There are only few experimental data that were obtained from tensile tests for whiskers~\cite{Kelly:1986} and from nanoindentation experiments.~\cite{Yoshida:2007}
Recently, calculations of ideal strength became of great interest because they represent an upper bound to the strength of any real crystal on the attainable stress.
The ideal strength may be calculated under special loading conditions: ideal strength in tension, e.g., for high symmetry directions of cubic crystals under uniaxial load, or ideal strength in shear, e.g., for common slip systems $<111>\{112\}$ and $<111>\{110\}$ in body-centered cubic (bcc) crystals.~\cite{crack,study:8}

The ideal strength is one of few mechanical properties that can be calculated from first principles. The ideal strength of refractory metals (such as Mo, Nb, V, and W), late transition metals (such as Cu, Pt and Au), elements crystallizing in the diamond structure (Si, Ge, and C) as well as some ordered alloys (TiAl and Ni$_{3}$Al) have been extensively  investigated.~\cite{study:1,study:2,study:3,study:4,study:5,study:6,study:7,study:8,study:9,Mo:1,stress,v:pp,Cerny:2010}
For example, Luo \emph{et al.}~\cite{Mo:1} and Nagasako \emph{et al.}~\cite{study:8} focused on the ideal tensile strength (ITS) of the bcc metals Mo, Nb, and V in the $<001>$ directions. Accordingly, in all of them the deformation starts along the Bain path, but branches away onto an orthorhombic path before the face-centered cubic (fcc) point is reached. In Mo, this orthorhombic distortion does not influence the ITS; however, in Nb and V, the branching occurs before the maximum stress on the tetragonal deformation curve is reached and hence causes a significant decrease in the ITS. The stress-strain relations and the corresponding theoretical tensile strengths of ordered $\gamma$-TiAl (\emph{L}$1_{0}$ structure) alloy exhibit strong anisotropy in different crystalline directions originating from the structural anisotropy of $\gamma$-TiAl.~\cite{study:6}
In spite of all these theoretical efforts, the first-principles description of the ITS in random solid solutions is rather limited. The only attempt~\cite{Li:2007} employed the virtual crystal approximation (VCA) to study the ITS of the binary Ti-V for concentrations of Ti equal to and higher than $30$ at.\%. The results showed the more Ti is present in the alloy the more the ITS is decreased.

Vanadium rich ternary V-Cr-Ti alloys are important candidate structure materials for the first-wall/blanket of future fusion reactors.~\cite{candidate:1,candidate:2} These V-based alloys exhibit excellent mechanical properties, decent thermal creep behavior, high thermal conductivity, good resistance to irradiation-induced swelling and damage, and long operating lifetime in the fusion environment.~\cite{candidate:1,candidate:2,candidate:3,candidate:4,candidate:5,candidate:6} Considerable efforts have been made to find optimal V-based alloy compositions that can endure the extreme environment of fusion-reactors. The available experimental data indicates that reasonable properties can be achieved by introducing a few percent Ti and Cr into the V matrix.~\cite{V:matrix} Consequently, the ternary V-Cr-Ti system has attracted broad interest, and in particular the compositions with $0-15$ at.\% Cr and $0-20$ at.\% Ti have been intensively investigated.~\cite{ternary:1,ternary:2,ternary:3,ternary:4,ternary:5} To 
the best of our knowledge, most experimental efforts on V-Cr-Ti alloys have been devoted to the ductile-brittle transition temperature before and after irradiation, swelling properties, and impact toughness as a function of Cr and Ti contents.

As a promising structure material for fusion reactors, V-based alloys must not only withstand radiation damage but should also keep intrinsic mechanical properties and structural strength. Thus it is necessary to study fundamental mechanical properties, such as elastic properties, which were systematically investigated for V$_{1-x-y}$Cr$_{x}$Ti$_{y}$ random alloys ($0\leq x\leq 0.1$ and $0\leq y \leq0.1$) in our former works.~\cite{Xiaoqing,xiaoqing:2011} The second-order elastic constants describe the mechanical properties of materials in the small deformation region, where the stress-strain relations are linear. The ideal strength describes the mechanical properties of the material beyond the elastic region, which is important during the alloy design.

In this work, we use the all-electron exact muffin-tin orbitals method (EMTO) in combination with the coherent-potential approximation (CPA) to investigate the ITS of elemental bcc V and Mo solids as well as Mo$_{0.9}$Tc$_{0.1}$ and V$_{1-x-y}$Cr$_{x}$Ti$_{y}$ ($0\leq x+y \leq0.1$) random alloys as a function of concentrations. The primary purpose of our work is to give an account of the application of the EMTO method to the calculation of ITS in bcc nonmagnetic metals and random alloys. Second, we aim to provide a consistent theoretical guide to further optimization of the composition of V-based alloys for structure material applications.

The structure of the manuscript is as follows. In Section~\ref{sec:computationalmethod}, we describe the computational tool and all important numerical details. The results are presented in Section~\ref{sec:resultsanddiscussion}. Here, first we assess the accuracy of our calculations by considering pure Mo and V and then we study the effects of the alloying elements on the ideal strength of Mo-Tc and V-Ti-Cr alloys.
We continue in Section~\ref{sec:discussion} with a discussion about the anisotropy of the attainable ideal strength and the trend of the computed ideal strength of Mo-Tc and the V-based alloys for the case of an isotropic Poisson contraction making use of band filling arguments, canonical band theory, and structural energy differences.
to explain the anisotropy of the attainable ideal strength and the trend of the computed ideal strength of Mo-Tc and the V-based alloys.

\section{\label{sec:computationalmethod}Computational method}
\subsection{Total energy calculation}
The first-principle method used in this work is based on density functional theory (DFT)~\cite{DFT} formulated within the Perdew-Burke-Ernzerhof (PBE) generalized gradient approximation for the exchange-correlation functional.~\cite{PBE} The Kohn-Sham equations were solved using the EMTO method.~\cite{EMTO:1,EMTO:2,EMTO:3} The problem of disorder was treated within the CPA and the total energy is computed via the full charge-density technique.~\cite{cpa:1,cpa:2,cpa:3,cpa:4}

The EMTO method is an improved screened Korringa-Kohn-Rostoker (KKR) method,~\cite{EMTO:1} where the full potential is represented by overlapping muffin-tin potential spheres. Inside these spheres, the potential is spherically symmetric and constant in between. By using overlapping spheres, one describes more accurately the exact crystal potential compared to conventional muffin-tin or non-overlapping methods. Further details about the EMTO method and its self-consistent implementation can be found in previous works.~\cite{EMTO:1,EMTO:2,EMTO:3,cpa:3,cpa:4} The accuracy of the EMTO method for the equation of state and elastic properties of metals and alloys has been demonstrated in a number of previous works.~\cite{cpa:2,EMTO:2,work:3,work:4,work:5,work:6,work:7,Xiaoqing}However, to our knowledge, no CPA-based alloy theory has previously been employed in the ab initio determination of the ideal strength of random solid solutions. This may be ascribed to the limited accuracy of the classical CPA-based 
electronic structure methods for systems with reduced crystal symmetry. In this respect, the present work represents a pioneering demonstration of the EMTO-CPA approach in ideal strength calculations for random alloys.

\subsection{Ideal tensile strength calculations for bcc crystals }

The principles of the response of bcc crystals to uniaxial loading were developed by Milstein \emph{et al.}~\cite{Milstein:1970,Milstein:1971,Milstein:1980,Milstein:1988}
The first step in ITS calculations is to compute the equilibrium lattice constant of the material in the ground state structure. The present alloys adopt the bcc structure. In the second step, an uniaxial tensile strain $\epsilon$ is applied along a specific crystalline direction which mimics a certain tensile stress $\sigma$. For each value of the strain, we relaxed the deformed structures, partially under constraints described below, to make sure that no internal forces remain in the crystal in directions perpendicular to the applied stress. From the above steps, we obtained energy versus strain curves and derived stress versus strain data. The first maximum on the stress-strain curve defines the ITS $\sigma_{\text{m}}$ for the selected strain path.

\begin{figure*}[hct]
\begin{center}
\begin{tabular}{c@{\qquad}c@{\qquad}c}
\resizebox{!}{4cm}{\includegraphics[clip]{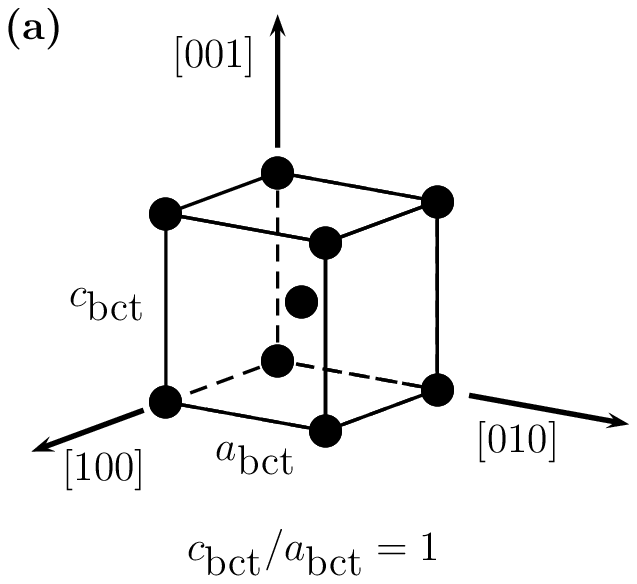}}%
&
\resizebox{!}{4cm}{\includegraphics[clip]{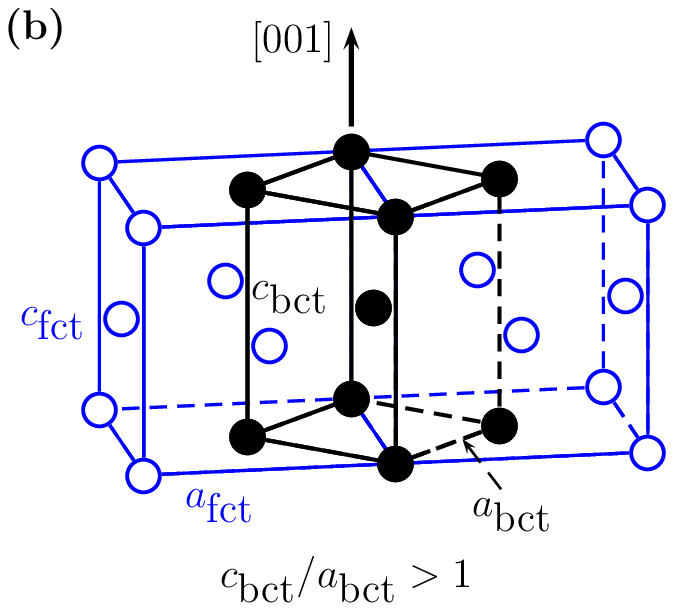}}%
\\
\end{tabular}
\caption{\label{fig:unitcell:100}(Color online) Illustration of the lattice distortion of the bcc structure due to an applied uniaxial stress in the $[001]$ direction. (a) The unit cell in the absence of strain is bcc (in the bct delineation $c_{\text{bct}}/a_{\text{bct}}=1$). For finite strain, the lattice symmetry is distorted to bct (b), and becomes fcc for $c_{\text{bct}}/a_{\text{bct}}=\sqrt{2}$. To describe the bifurcation from the primary tetragonal to the secondary orthorhombic strain path, the fct reference frame is used.}
\end{center}
\end{figure*}

The stress $\sigma$ is given by~\cite{stress}
\begin{equation}
\sigma(\epsilon)=\frac{1}{\Omega(\epsilon)}\frac{\partial\emph{E}}{\partial\epsilon},
\label{eq:stress}
\end{equation}
where $\emph{E}$ is the total energy per atom and $\Omega(\epsilon)$ is the volume per atom at a given tensile strain. $\epsilon$ is the strain of the simulation cell in the direction of the applied uniaxial force $\hat{\textbf{F}}$ and is defined as
\begin{equation}
\epsilon=\frac{l_{\parallel}-l_{0}}{l_{0}},
\label{eq:strain}
\end{equation}
where $l_{\parallel}$ and $l_{0}$ denote the length of the cell parallel to $\hat{\textbf{F}}$ in the final state and in the initial state (without any force), respectively. The initial state corresponds to the equilibrium bcc structure with energy $E_0$.
We define the \emph{uniaxial strain energy} $\Delta E(l_{\parallel},\hat{\textbf{F}})$ following Ref.~\onlinecite{Ozolins:1998} as the total energy change upon deforming the material in the direction along the applied force, and relaxing with respect to the dimensions in the plane perpendicular to $\hat{\textbf{F}}$, viz.
\begin{equation}
\Delta\emph{E}(l_{\parallel},\hat{\textbf{F}})= \min_{\mathbf{a}_{1},\mathbf{a}_{2}} E(\mathbf{a}_{1},\mathbf{a}_{2},l_{\parallel}) - E_0.
\label{eq:energy}
\end{equation}
The minimization is done with respect to the pair of unit cell vectors $\{\mathbf{a}_{1},\mathbf{a}_{2}\}\perp\hat{\textbf{F}}$.
\begin{figure*}[htbc]
\begin{center}
\begin{tabular}{c@{\qquad}c}
\resizebox{!}{4cm}{\includegraphics[clip]{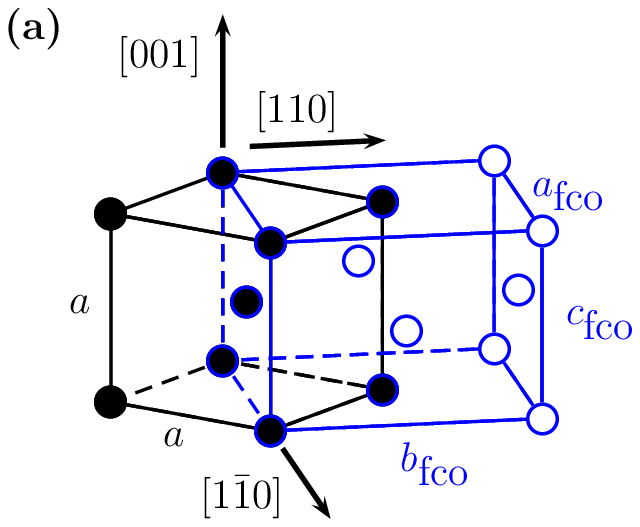}}
&
\resizebox{!}{4cm}{\includegraphics[clip]{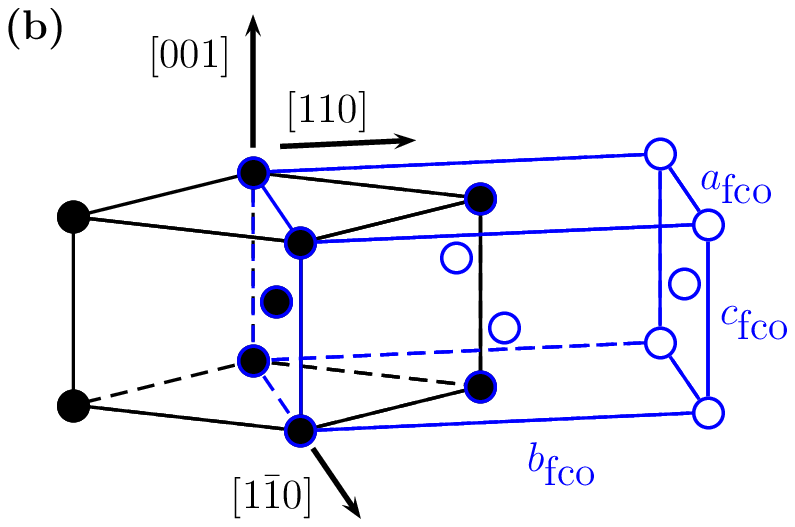}}
\end{tabular}
\caption{\label{fig:unitcell:110}(Color online) Illustration of the lattice distortion of the bcc structure due to an applied uniaxial stress in the $[110]$ direction. (a) and (b) show the undistorted lattice and the distorted lattice, respectively. Both the fco unit cell used in the computation and the (undistorted and distorted) bcc cell are sketched.}
\end{center}
\end{figure*}

In this work, we computed the ITS in the $<001>$, $<111>$ and $<110>$ directions of bcc elemental solids and alloys. Each direction breaks the guiding symmetry of the bcc structure in a different way.
First, we chose the $[001]$ axis to be the direction of the applied force for $<001>$ ITS calculations. The symmetry of the bcc lattice is reduced to the body-centered tetragonal (bct) one on the primary deformation path, see Fig.~\ref{fig:unitcell:100}. Accordingly, we may rewrite Eq.~\eqref{eq:energy} with the in-plane lattice constant $a_{\text{bct}}\equiv a_1=a_2$, $a_i=|\mathbf{a}_i|$, and $\mathbf{a}_1\cdot\mathbf{a}_2=0$,
\begin{equation}
\Delta\emph{E}^{\text{\,prim}}(c_{\text{bct}},[001])= \min_{a_{\text{bct}}} E(a_{\text{bct}},c_{\text{bct}}) - E_0,
\label{eq:energy:001}
\end{equation}
with $c_{\text{bct}}\equiv l_{\parallel}$ denoting the unit cell length along $[001]$.
Increasing the axial ratio $c_{\text{bct}}/a_{\text{bct}}$ from 1 to $\sqrt{2}$ transforms the bcc lattice into the fcc lattice while the crystal remains bct during the transformation. This transformation corresponds to the Bain transformation.~\cite{Bain:1924} 

It is known from previous calculations for V~\cite{Li:2007,study:8} as opposed to Mo,~\cite{Mo:1} that a bifurcation to a secondary orthorhombic strain path occurs before the ITS along the primary tetragonal strain path is passed leading to a significant reduction of the ITS.
The observed orthorhombic branching is defined with respect to the face-centered tetragonal (fct) reference frame of the primary deformation path, see Fig.~\ref{fig:unitcell:100}, while the branching would correspond to a monoclinic deformation in the bct reference frame. We used the fct reference frame to describe the bifurcation in this work. Accordingly, we denote the uniaxial strain energy of the secondary deformation path by
\begin{equation}
\Delta\emph{E}^{\text{\,secon}}(c_{\text{fco}},[001])= \min_{a_{\text{fco}},b_{\text{fco}}} E(a_{\text{fco}},b_{\text{fco}},c_{\text{fco}}) - E_0,
\label{eq:energy:001orth}
\end{equation}
with $a_{\text{fco}}$, $b_{\text{fco}}$, and $c_{\text{fco}}$, $c_{\text{fco}}\equiv c_{\text{bct}}$, denoting the lattice parameters of the face-centered orthorhombic (fco) lattice.

In the actual calculation of the stress-strain relation we selected a set of $c_{\text{bct}}$ ($c_{\text{fco}}$) values equidistantly separated by $2.2\%$ of the bcc equilibrium lattice parameter corresponding to different strains $\epsilon$. For each strain value, we computed the total energy on a dense mesh of $a_{\text{bct}}$ ($a_{\text{fco}}$ and $b_{\text{fco}}$) values to find the minimum according to Eq.~\eqref{eq:energy:001} (Eq.~\eqref{eq:energy:001orth}). The total energies as well as the volume of each state of strain were fitted to polynomial fit functions and differentiated to obtain the stress as a function of strain. The maximum strain, $\epsilon_\text{m}$, is then the ITS at the maximum stress, $\sigma_{\text{m}}$, on a given strain path. The choice of the order of the polynomial fit functions was crosschecked to avoid any substantial influence on the values of $\sigma_\text{m}$ and of $\epsilon_{\text{m}}$. For consistency, we employed the same fit function for all types of distortion, i.e., 
along the $<001>$, $<111>$, and $<110>$ directions.

\begin{figure*}[hbtc]
\begin{center}
\begin{tabular}{c@{\qquad}c@{\qquad}c}
\resizebox{!}{4cm}{\includegraphics[clip]{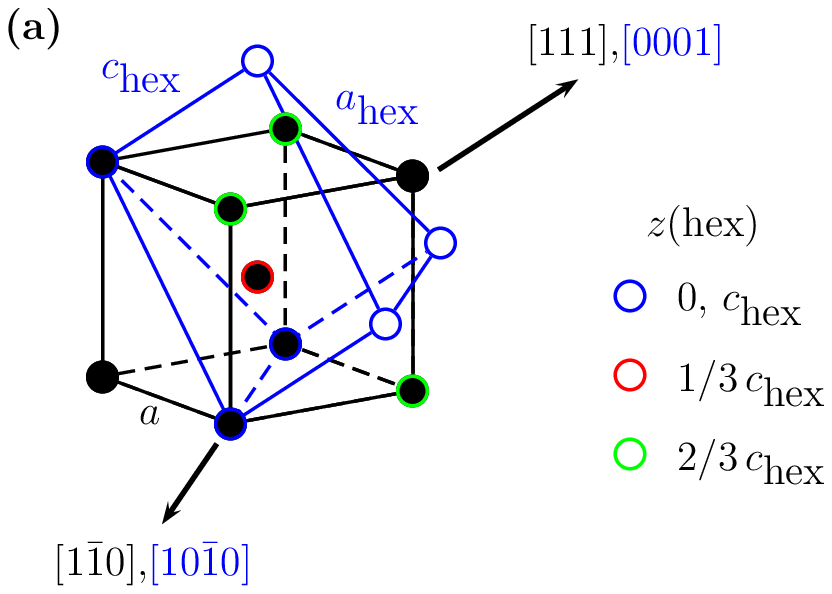}}%
&
\resizebox{!}{4cm}{\includegraphics[clip]{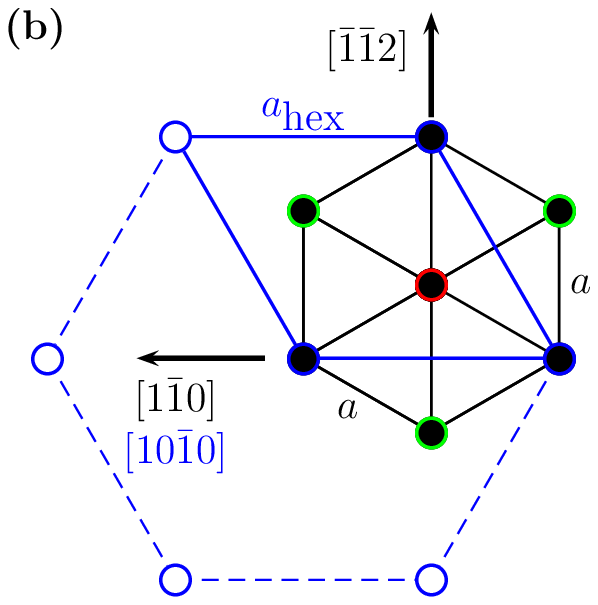}}%
&
\resizebox{!}{4cm}{\includegraphics[clip]{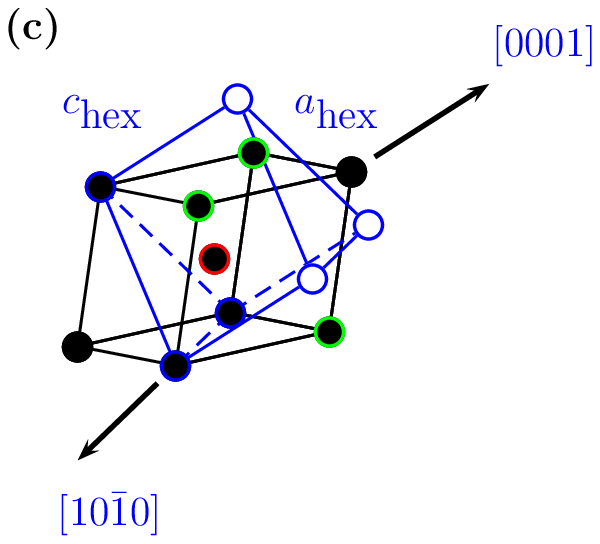}}%
\\
\end{tabular}
\caption{\label{fig:unitcell:111}(Color online) Illustration of the lattice distortion of the bcc structure due to an applied uniaxial stress in the $[111]$ direction. (a) shows the undistorted bcc lattice with hexagonal delineation (a wedge representing one sixth of the conventional hex unit is drawn), the axial ratio is $c_{\textrm{hex}}/a_{\textrm{hex}}=\sqrt{3/8}$. The undistorted unit cell projected along $[111]$ ($[0001]$) is depicted in (b) to show more clearly the full symmetry of the hex lattice. A finite strain, (c), lowers the symmetry of the bcc lattice to a trigonal one.}
\end{center}
\end{figure*}

Straining the bcc structure along the $[110]$ axis reduces the lattice symmetry to fco guiding symmetry.  In the absence of strain, the fco lattice parameters fulfill $a_{\text{fco}}=b_{\text{fco}}=\sqrt{2}c_{\text{fco}}$, see Fig.~\ref{fig:unitcell:110} for a detailed illustration. Here, strain is applied to $b_{\text{fco}}\equiv l_{\parallel}$. Assuming isotropic Poisson contraction, $a_{\text{fco}}$ and $c_{\text{fco}}$ are relaxed keeping $c_{\text{fco}}/a_{\text{fco}}$ fixed to the initial value of $\sqrt{2}$. This assumption was used to make the computations feasible. Previous investigations for V identified the $[110]$ direction as the strongest one among $[001]$, $[110]$, and $[111]$.~\cite{paw:2} Thus, the $[110]$ direction may be considered to be of least interest in ITS calculations. We may hence write for the uniaxial strain energy of the $[110]$ distortion,
\begin{equation}
\Delta\emph{E}(b_{\text{fco}},[110])= \min_{a_{\text{fco}}} E(a_{\text{fco}},b_{\text{fco}}) - E_0.
\label{eq:energy:110}
\end{equation}

If an uniaxial strain is applied along the $[111]$ direction parallel to the body diagonal of the bcc structure, the symmetry of the lattice is reduced to trigonal symmetry. The distorted lattice can be equivalently described by a hexagonal (hex) lattice or a rhombohedral lattice. Here, we chose the hex lattice, as sketched in  Fig.~\ref{fig:unitcell:111}. The $[0001]$ axis and the $[10\bar{1}0]$ axis of the hex lattice are oriented parallel to the $[111]$ axis and the $[1\bar{1}0]$ axis of the bcc lattice, respectively. In the absence of strain, the lattice parameters of the hex unit cell and the bcc cell are related by $a_{\text{hex}}=\sqrt{2}a_{\text{bcc}}$ and $c_{\text{hex}}=\sqrt{3/4}a_{\text{bcc}}$, where $a_{\text{hex}}$ denotes the in-plane lattice parameter of the hexagonal basal plane (the distance between two atoms of the smallest equal-sided triangle). The out-of-plane lattice parameter, $c_{\text{hex}}$, is oriented parallel to the threefold stacking axis. Bcc (111) planes are stacked along $[
0001]$ (ABCABC stacking) and two successive planes are displaced by 1/3 of $c_{\text{hex}}$ (cf.\! Fig.~\ref{fig:unitcell:111}).
For each applied strain, interatomic distances in the $(0001)$ planes are relaxed while the in-plane symmetry is preserved. Thus, we assume isotropic Poisson contraction for the $[111]$ type of ITS calculations noting that a branching away from trigonal symmetry has not been reported in previous investigations for V.~\cite{paw:2}

The uniaxial strain energy relaxed with respect to $a_{\text{hex}}\equiv a_1=a_2$ becomes,
\begin{equation}
\Delta\emph{E}(c_{\text{hex}},[111])= \min_{a_{\text{hex}}} E(a_{\text{hex}},c_{\text{hex}}) - E_0.
\label{eq:energy:111}
\end{equation}
When the hexagonal axial ratio $c_{\text{hex}}/a_{\text{hex}}$ is identical to $\sqrt{3/2}$ the strained hexagonal lattice coincides with the simple cubic (sc) lattice. Similarly, when $c_{\text{hex}}/a_{\text{hex}}$ =$\sqrt{3}$ we recover the fcc lattice.

\section{\label{sec:resultsanddiscussion}Results}
\subsection{\label{sec:equilibrium}Equilibrium volume}
The theoretical equilibrium lattice parameter of bcc Mo is $3.165${\AA}, which agrees well with the experimental data~\cite{Kittel:1986} $3.141${\AA}. When $10$ at.\% Tc is added into the Mo matrix, the lattice constants decreases from $3.165${\AA} to $3.156${\AA}. This trend can be understood by the smaller atomic radius of hexagonal close packed Tc ($1.36$ {\AA}) as compared to that of bcc Mo ($1.40$ {\AA}).~\cite{Kittel:1986}

Theoretical equilibrium lattice parameters and elastic properties of V and disordered V-Cr-Ti alloys were investigated with EMTO and the CPA in our previous paper, see Ref.~\onlinecite{Xiaoqing}. Since our employed lattice parameters are identical to these earlier results, we refer the reader to this reference for a detailed account of the numerical data.
Here we only restate the most important findings necessary to follow the ongoing discussion.

For pure vanadium, the calculated equilibrium lattice parameter is $2.998${\AA}, in good agreement with the corresponding experimental value~\cite{lattice} $3.03$ {\AA} at low temperature. For V alloyed with Ti and Cr, we found that the theoretical lattice constants increase (decrease) with increasing Ti (Cr) concentrations, for instance, the lattice constant of the alloy with 10\% Cr is $2.981$ {\AA}, but for the alloy with 10\% Ti, the lattice constant is $3.021${\AA}. Our theoretical results confirmed the experimental trend.~\cite{lattice:alloy}

Our results showed that the EMTO method with the CPA can be used to describe the elastic properties of V-based alloys and gave the accuracy and available results in the small deformation region. Since the ideal strength describes the mechanical properties of the material beyond the linear stress-strain response, the ideal strength is very important during the design process of alloys. Thus, based on the equilibrium lattices constants and energies that we obtained in our former calculations,~\cite{Xiaoqing} we calculated the ideal tensile strength of pure V, and investigated the ideal tensile strength of bcc nonmagnetic V$_{1-x-y}$Cr$_{x}$Ti$_{y}$ random alloys as a function of Cr and Ti ($0\leq x+y \leq0.1$) concentrations using EMTO with the CPA.

\subsection{\label{sec:strengthV}Ideal strength of pure vanadium and molybdenum}
To assess reliability of our computational approach, we first performed the simulation of a tensile test in V for uniaxial loadings along the $[001]$, $[110]$ and $[111]$ directions.  A branching to a secondary orthorhombic path under uniaxial loading along $[001]$ was predicted for V~\cite{Li:2007,study:8}. The branching point occurs before the maximum stress on the primary tetragonal deformation path is reached. Here, using our computational approach, we investigated this bifurcation for V with the aim to reveal its impact on the attainable ITS.

\begin{figure}[hct]
\begin{center}
\resizebox{0.9\columnwidth}{!}{\includegraphics[clip]{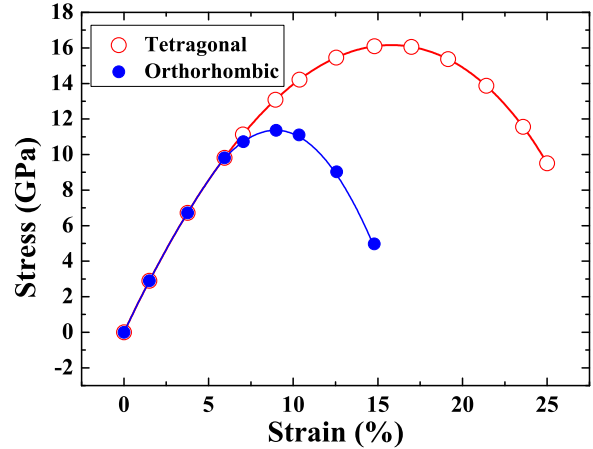}}
\caption{\label{fig:orthorhombic}(Color online) The stress of bcc V under $[001]$ tension as a function of the applied strain. Open symbols and closed symbols refer to the primary tetragonal deformation path and to the secondary orthorhombic deformation path, respectively.}
\end{center}
\end{figure}

The corresponding stresses as a function of strain along the $[001]$ direction are displayed in Fig.~\ref{fig:orthorhombic}. Along the tetragonal deformation path, stress increases with increasing strain up to a maximum of $\sigma_\text{m}=16.1$ GPa at a strain of $\epsilon_{\text{m}}=16.0\%$. The significantly lower ideal stress and
strain on the orthorhombic deformation path are clearly different from those corresponding to the tetragonal deformation path. The stress corresponding to the secondary orthorhombic path reaches a maximum of $11.6$ GPa at $\epsilon_{\text{m}}=9.0\%$. Hence the ITS of V is limited by the bifurcation to the secondary orthorhombic path in the $[001]$ direction, in line with the previous observations~\cite{study:8,Li:2007}.

\begin{table*}[ht]
\centering
\caption{\label{table:one}The ideal tensile strength $\sigma_{\text{m}}$ and the corresponding strain $\epsilon_{\text{m}}$ for the tetragonal (V and Mo) and orthorhombic (V) deformation paths in the $[001]$ direction.}
\begin{ruledtabular}
\begin{tabular}{cccccc}
\multirow{2}{*}{system} & \multirow{2}{*}{method} & \multicolumn{2}{c}{Tetragonal} & \multicolumn{2}{c}{Orthorhombic} \\
\cline{3-6}
   & & $\sigma_{\text{m}}$ (GPa) & $\epsilon_{\text{m}} (\%)$ & $\sigma_{\text{m}}$ (GPa) & $\epsilon_{\text{m}} (\%)$  \\
 \hline
    \multirow{6}{*}{V} & This work & 16.1 & 16.0 &11.6& 9.0  \\
      &  PAW Ref.~\onlinecite{paw:2} & 19.1 & 18.0  \\
      &  PAW Ref.~\onlinecite{study:8} & 17.8 & 17.0& 11.5 &10.0  \\
      &  PAW Ref.~\onlinecite{paw:3} & 18.9 & 18.2 &  & \\
      &  PP Ref.~\onlinecite{Li:2007}\footnotemark[1]  & 17& 16.0 &14 &12.0  \\
      &  PP Ref.~\onlinecite{v:pp}  & 12.2 & 8.3  \\
\hline
      \multirow{3}{*}{Mo} & This work & 26.7 & 12.0 \\
   &  FP-LAPW Ref.~\onlinecite{Mo:2} & 25.4 & 10.0 \\
   &  PP Ref.~\onlinecite{Mo:1} & 28.8 & 13.0  \\
\end{tabular}
\end{ruledtabular}
\footnotetext{values estimated from figure}
\end{table*}

\begin{figure}[htb]
\begin{center}
\resizebox{0.8\columnwidth}{!}{\includegraphics[clip]{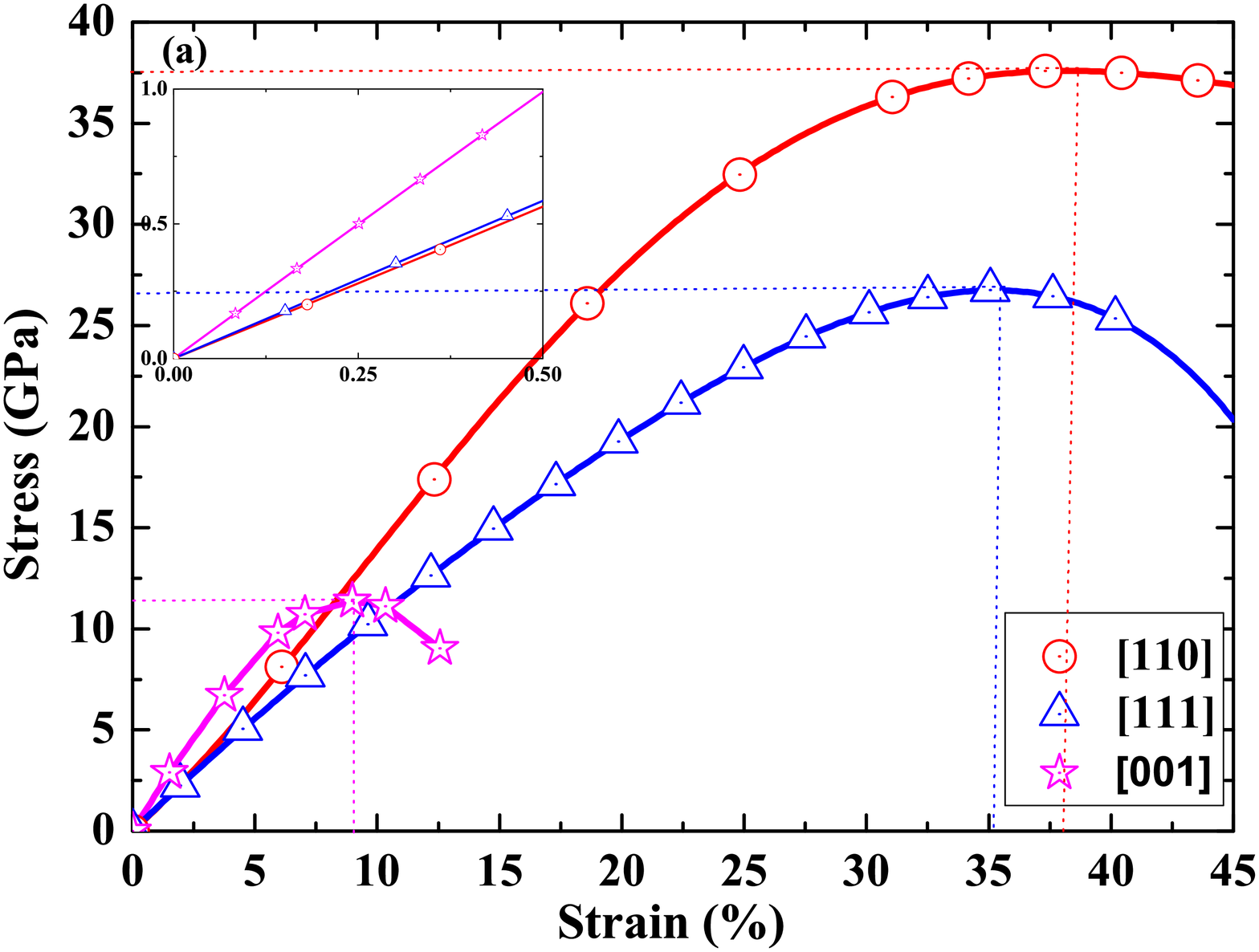}}
\newline
\resizebox{0.8\columnwidth}{!}{\includegraphics[clip]{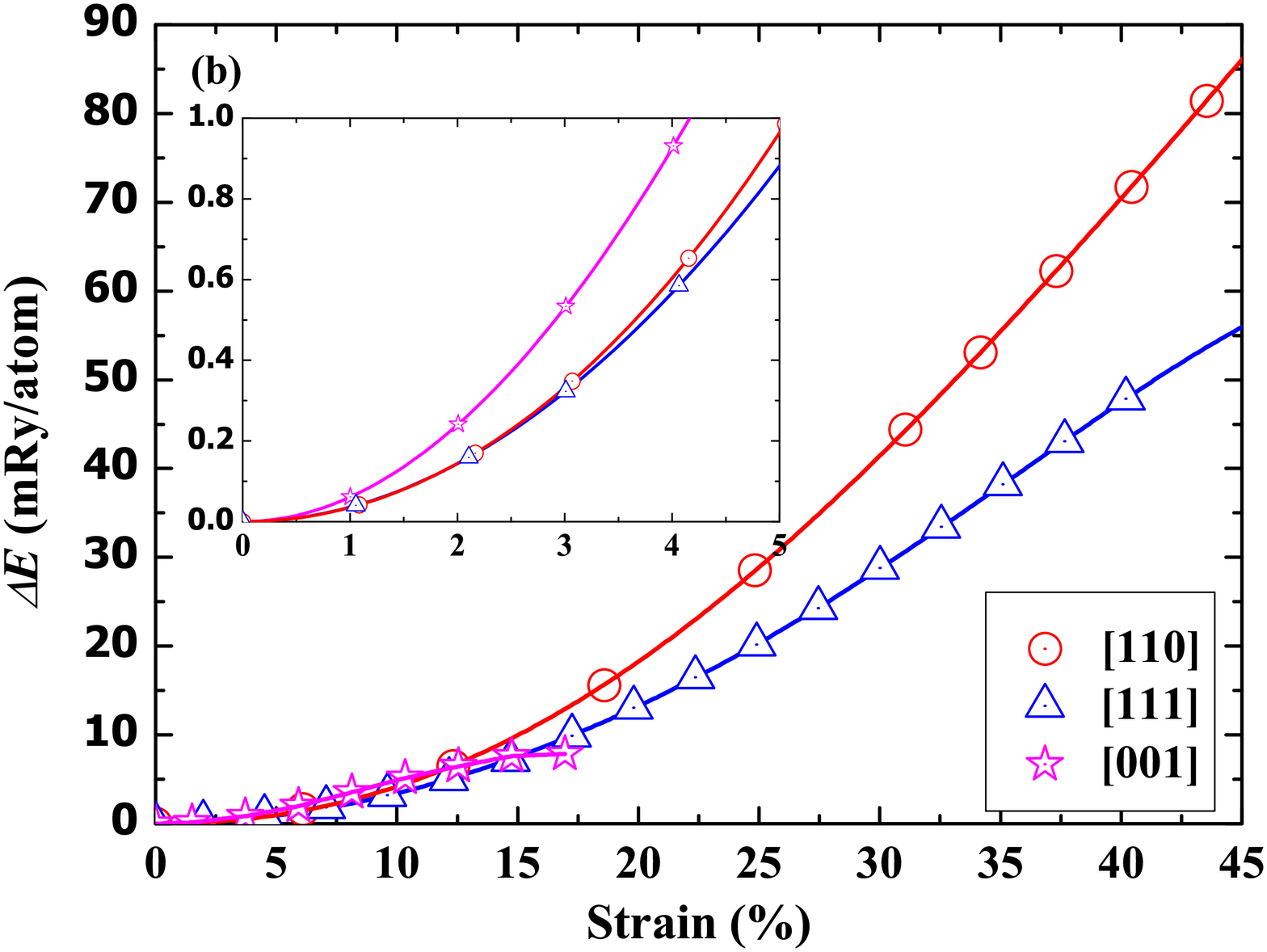}}
\newline
\caption{\label{fig:vstress}(Color online) The stress (a) and energy (b) of pure V in $[001]$ (secondary deformation path), $[111]$ and $[110]$ directions as a function of strain. The insets show the trends of ideal strength and energy in the smaller strain region.}
\end{center}
\end{figure}

Figure~\ref{fig:vstress} displays the stresses as a function of strain along all investigated directions for V. Compared to the $[001]$ direction,  the ideal stresses are much higher in the $[111]$ direction and the $[110]$ direction for which the stress reaches a maximum of $26.7$ GPa at $\epsilon_{\text{m}}=35.2\%$ and $37.6$ GPa at $\epsilon_{\text{m}}=38.3\%$, respectively.

In the inset of Fig.~\ref{fig:vstress}(a), we also show stress as a function of strain in the small deformation region ($\epsilon\leq0.5\%$). These strain-stress relations are linear and follow Hooke's law, $\sigma=E\cdot\epsilon$,
where $E$ is Young's modulus, which depends on the direction of the applied force.~\cite{Grimvall:1999} For further use, the explicit expression of $E$ in terms the cubic elastic constants ($C_{11}$, $C_{12}$, and $C_{44}$) for an uniaxial stress in $<001>$ directions ($E_{<001>}$) and in  $<111>$ directions ($E_{<111>}$) are given by
\begin{eqnarray}
E_{<001>}&=&6C'\cdot\frac{B}{C_{11}+C_{12}} \\
E_{<111>}&=&3C_{44}\cdot\frac{1}{1+\frac{C_{44}}{3B}},
\end{eqnarray}
where $C'=(C_{11}-C_{12})/2$ and $B=(C_{11}+2C_{12})/3$ are the tetragonal shear elastic constant and the bulk modulus, respectively. According to Fig.~\ref{fig:vstress}(a) (inset), the $[001]$ direction exhibits the largest Young's modulus, which means that the $[001]$ direction is the strongest direction among the three selected ones in the small deformation region, while the Young's moduli in the $[110]$ and $[111]$ directions are smaller and degenerate. Indeed, using the theoretical elastic parameters of bcc V~\cite{Xiaoqing} we obtain $E_{<001>}$=$205.3$ GPa and $E_{<111>}$=$101.4$ GPa. These values are in good agreement with those derived directly from Fig.~\ref{fig:vstress}(a) (approximately 200\,GPa and 110\,GPa for $E_{<001>}$ and  $E_{<111>}$, respectively). This behavior at small strains is distinct from the one for the larger, non-linear deformation region, where $[110]$  ultimately replaces $[001]$ as the strongest direction.

The ideal tensile strength $\sigma_{\text{m}}$ corresponding to the strain $\epsilon_{\text{m}}$ from our and other calculations in the $[001]$ direction and in the other two directions are listed in Table~\ref{table:one} and Table~\ref{table:aa}, respectively. In Table~\ref{table:one}, for comparison we included results obtained for both primary tetragonal and secondary orthorhombic path, as many of the previous calculations were performed only for the prior one. From Table~\ref{table:one}, it can be seen that for the tetragonal deformation path, projector-augmented-wave (PAW/) works~\cite{study:8,paw:2,paw:3} reported similar ideal strength values, such as Nagasako \emph{et al.}~\cite{study:8} gave a value of $\sigma_{\text{m}}$=$17.8$ GPa at $\epsilon_{\text{m}}$=$17.0$\% and Liu \emph{et al.} reported a value of $\sigma_{\text{m}}$=$19.1$ GPa at $\epsilon_{\text{m}}$=$18.0$\%.
Our EMTO results for stress and strain are slightly smaller than the PAW values however still in good agreement with them.
Li \emph{et al.}~\cite{Li:2007} tabulated a value of $\sigma_{\text{m}}$=$17$ GPa at $\epsilon_{\text{m}}$=$16.0$\% using pseudo potentials (PP). Our results are in agreement with their values.
Krenn \emph{et al.} obtained an ITS value of $12.2$ GPa at an attainable strain of $8.3$\%, which is considerably lower compared with all other theoretical results. The difference may be due to their estimation of the ideal stress assuming a sinusoidal stress-strain relation on the basis of Frenkel's~\cite{Frenkel:1926} and Orowan's~\cite{Orowon:1949} works. Compared to the tetragonal deformation path, the ITS and the maximum strain are significantly reduced along the orthorhombic path in accordance with previous investigations. The ITS of V decreases by about $4$ GPa, which is in line with the $6$ GPa and the $3$ GPa decrease as reported in Ref.~\onlinecite{study:8} (PAW) and Ref.~\onlinecite{Li:2007} (PP), respectively.

From Table~\ref{table:aa} we can see that our computed maximum stress for the $[110]$ direction is somewhat larger than the only available theoretical data, Ref.~\onlinecite{paw:2}. This may be attributed to our constraint relaxation (fixed lattice parameter ratio $c_{\text{fco}}/a_{\text{fco}}=\sqrt{2}$). If this constraint is released as done in Ref.~\onlinecite{paw:2}, the total energy for each lattice distortion lowers and hence the uniaxial strain-energy curve may be shallower compared to a constraint calculation.

Comparing the ITSs along the $[001]$, $[111]$ and $[110]$ directions, we identify the $[001]$ direction as the weakest one. Former ab initio approaches~\cite{inherent:property,stress,Mo:1,liu,study:2,sob:1998,Cerny:2007,Cerny:2010} also identified the $[001]$ direction as the weakest direction in tension for elements crystallizing in the bcc structure.
The ITSs of V along the $[111]$ and $[110]$ directions are clearly larger than the one along $[001]$, the ideal strength along the $[110]$ direction exhibiting the largest value. \v{C}ern\'{y} and Pokluda~\cite{Cerny:2007,Cerny:2010} investigated various bcc transition metals and also predicted the largest stresses for the $[110]$ direction.

\begin{table}[htbpc]
\caption{\label{table:aa} Comparison between the present and former (Ref.~\onlinecite{paw:2}) ideal tensile strength $\sigma_{\text{m}}$ and the corresponding strain $\epsilon_{\text{m}}$ of bcc V calculated in the $[111]$ and $[110]$ directions.}
\begin{ruledtabular}
\begin{tabular}{cccccc}
\multirow{2}{*}{element} & \multirow{2}{*}{method} & \multicolumn{2}{c}{direction/$[111]$} & \multicolumn{2}{c}{direction/$[110]$}\\
\cline{3-6}
   & & $\sigma_{\text{m}}$ (GPa) & $\epsilon_{\text{m}} (\%)$ & $\sigma_{\text{m}}$ (GPa) & $\epsilon_{\text{m}} (\%)$  \\
   \hline
  \multirow{2}{*}{V} & This work &26.7& 35.2& 37.6 & 38.3    \\
      & PAW Ref.~\onlinecite{paw:2}  &31.0&36.0 & 32.8 & 42.0\\
     \end{tabular}
\end{ruledtabular}
\end{table}

To further assess the accuracy of our computational approach, we calculated the ITS of bcc molybdenum.
Since the $[001]$ direction was already identified to be the weakest direction of bcc Mo and no branching to a secondary deformation path occurs before the ITS on the primary deformation path is reached,~\cite{Mo:1} here we concentrated on the primary deformation path along the $[001]$ direction only.
The ideal strength of bcc Mo obtained in this work and the previously published results are listed in Table~\ref{table:one}. One can see that the present results are very close to the other theoretical evaluations.

Based on the above findings, we conclude that our theoretical tool is able to describe the ideal strength of the V and Mo systems with sufficiently high accuracy and thus can be further employed to study V- and Mo- based alloys.

\subsection{Ideal strength of Mo-Tc and V-based alloys}

\begin{figure}[htb]
 \resizebox{0.9\columnwidth}{!}{\includegraphics[clip]{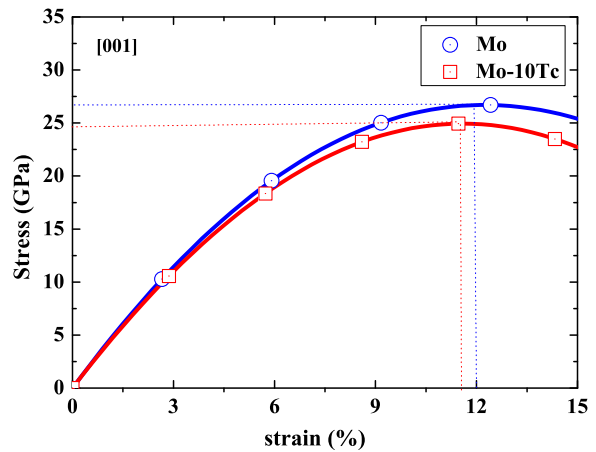}}%
\caption{\label{fig:Mo} (Color online) The ideal tensile strengths of Mo single crystal and the Mo-10Tc alloy as a function of strain in the $[001]$ direction.}
\end{figure}

Next, we investigated the alloying effect of Tc on the ideal strength of Mo in the concentration range up to $10$ at.\% Tc along the $[001]$ direction.
We took into account the possibility of a branching from the primary bct deformation path to the secondary orthorhombic deformation path for the present Mo-Tc-based alloys. Based on these calculations, we can exclude a bifurcation from the primary bct deformation before the ideal strength is reached, i.e., the Mo-10Tc alloy fails by cleavage.
The stress-strain dependence for Mo-Tc alloys is displayed in Fig.~\ref{fig:Mo} and the corresponding ITS data are listed in Table~\ref{table:two}.
From Fig.~\ref{fig:Mo}, we identify $\sigma_{\text{m}}=24.9$ GPa at $\epsilon$=$11.6\%$ for the Mo-10Tc alloy. Compared to pure Mo, Tc decreases the ideal strength. An explanation of the obtained alloying effect on the ITS of bcc Mo will be given in Section~\ref{sec:discussion}.

\begin{figure}[h!tb]
\begin{center}
\resizebox{0.8\columnwidth}{!}{\includegraphics[clip]{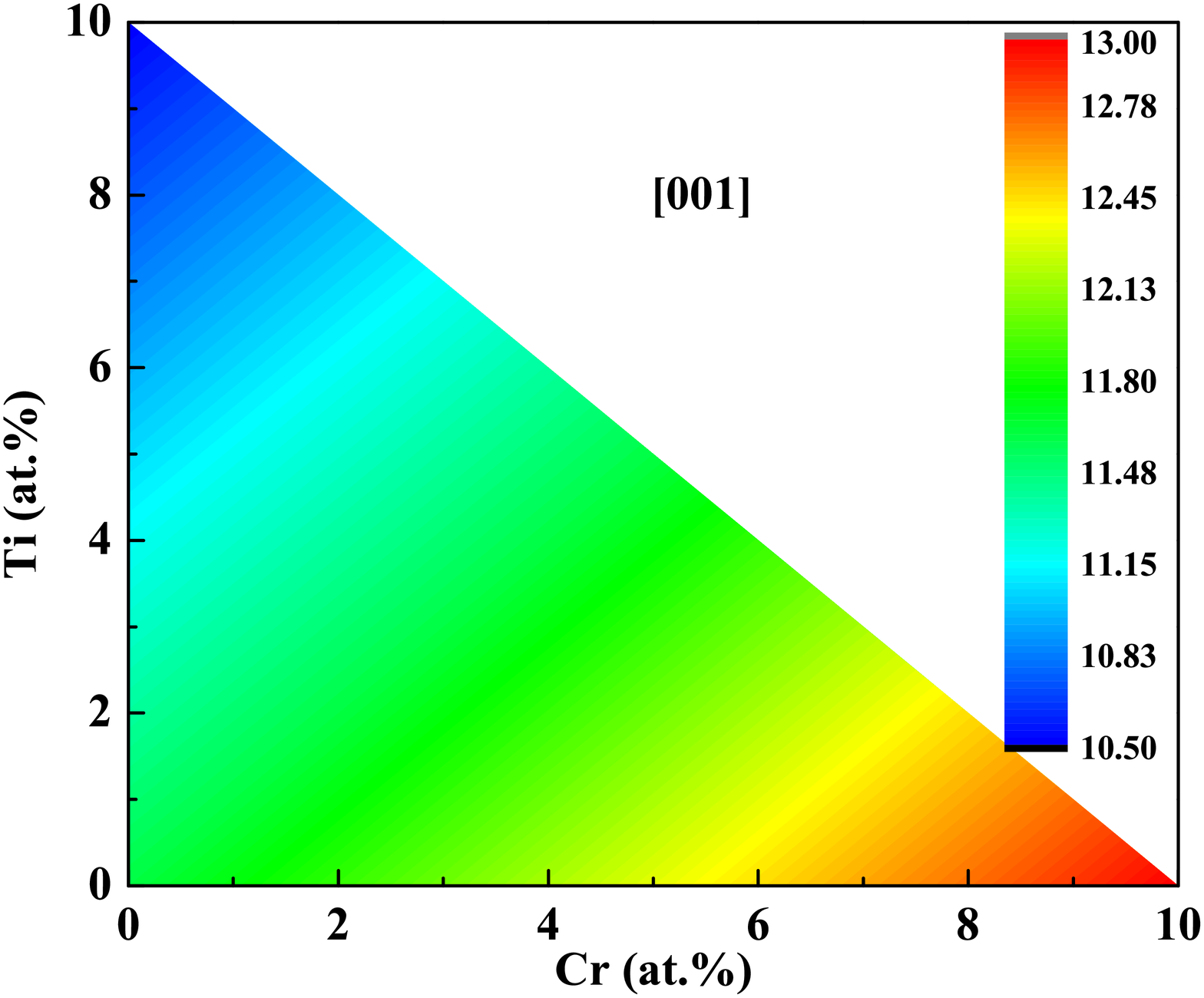}}%
\newline
\resizebox{0.8\columnwidth}{!}{\includegraphics[clip]{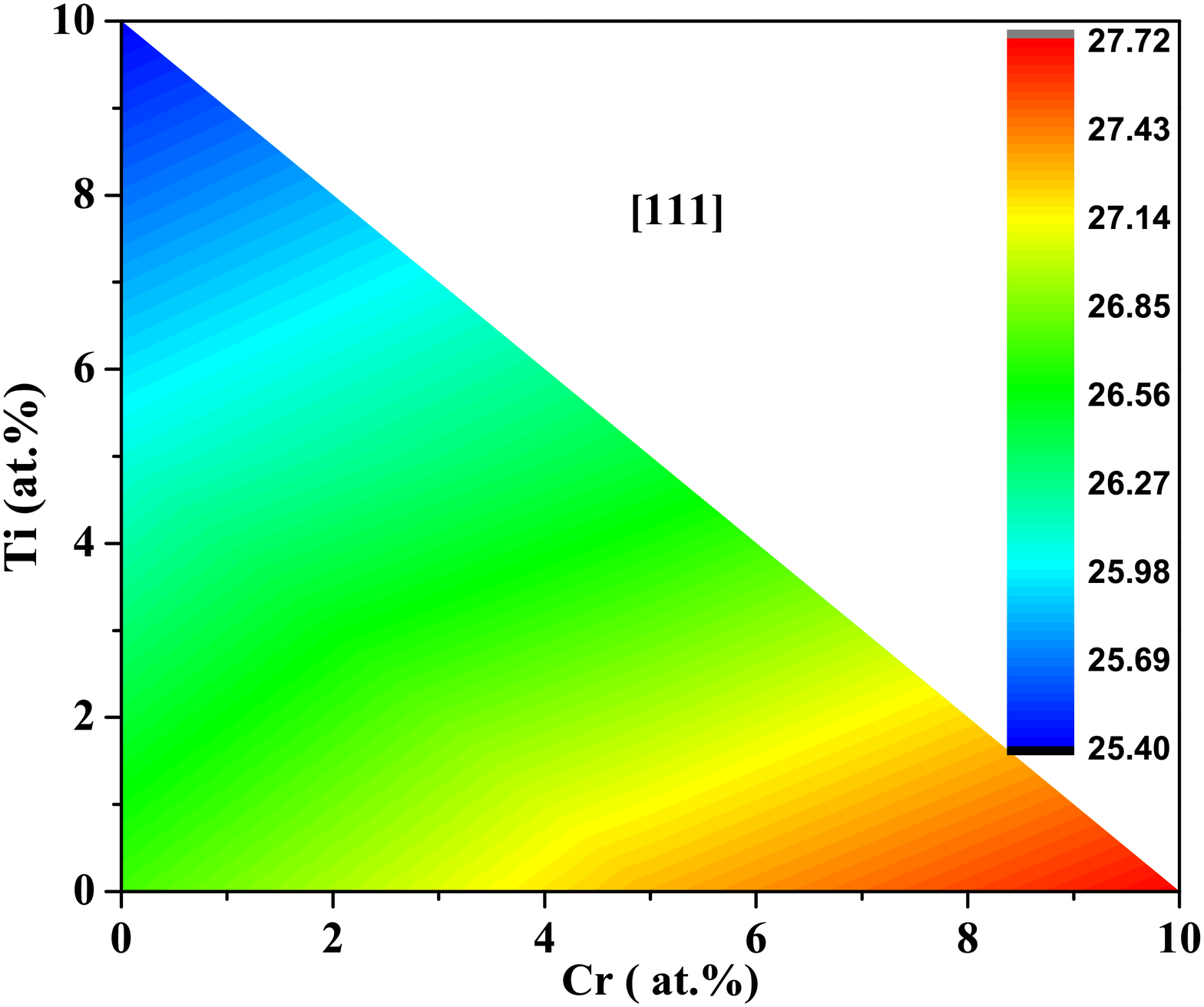}}%
\newline
\resizebox{0.8\columnwidth}{!}{\includegraphics[clip]{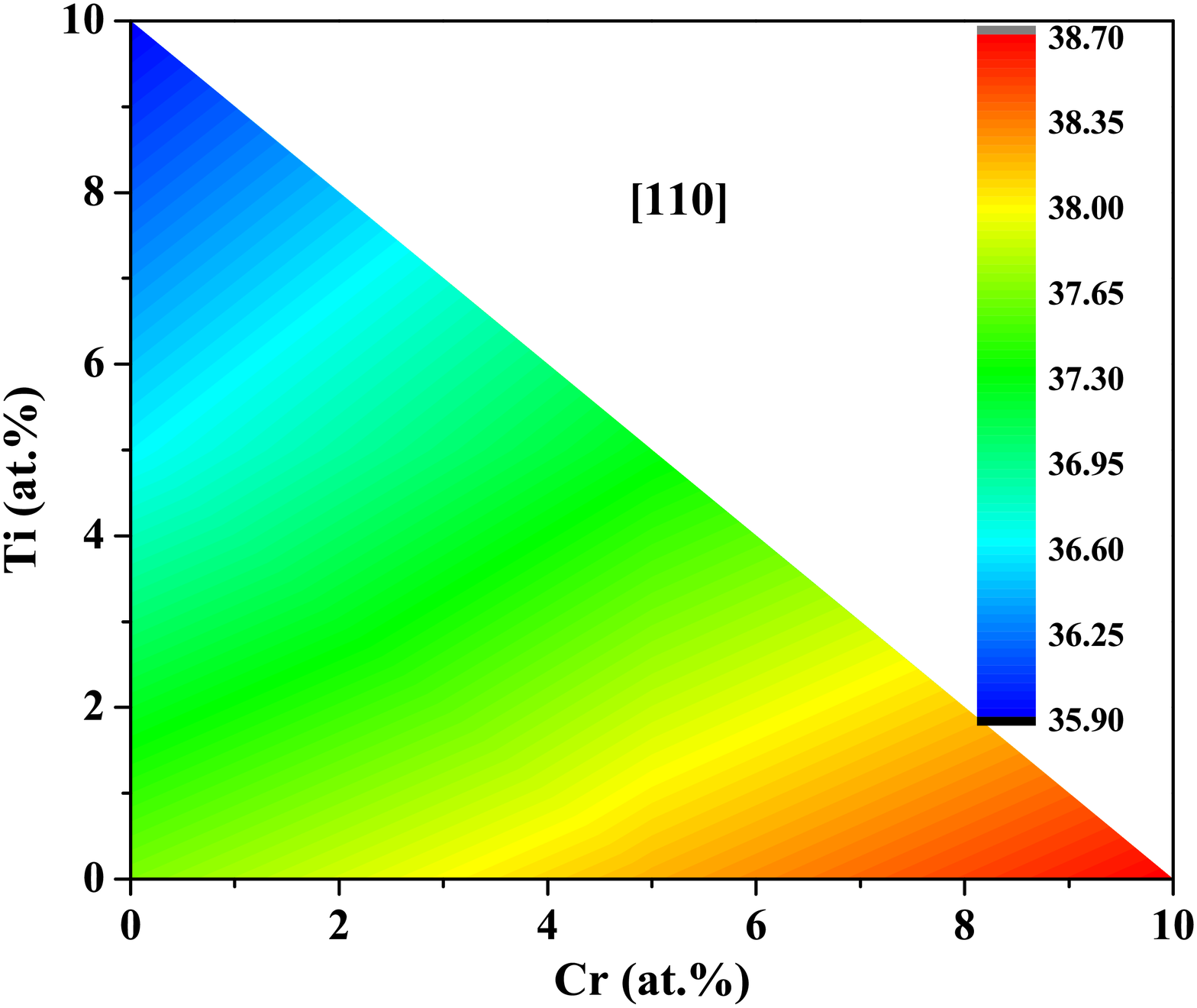}}%
\caption{\label{fig:alloys} (Color online) The ideal strength of V-based alloys in $[001]$, $[111]$ and $[110]$ directions as a function of Cr and Ti. All stress values are in GPa. The $[001]$ stress corresponds to the secondary (orthorhombic) path (i.e. fully relaxed structure within the plane perpendicular to the strain), whereas the $[111]$ and $[110]$ stresses correspond to isotropic Poisson contraction.}
\end{center}
\end{figure}

In the following we turn to the V-based alloys and investigated the effect of the alloying elements Cr and Ti on the ITS of bcc V. The total concentration of both solutes was varied in the range from 0 to 10\%.

Figure~\ref{fig:alloys} shows the composition dependence of the ideal strength of ternary bcc V-Cr-Ti random alloys along the $[001]$, $[111]$, and $[110]$ directions and the corresponding numerical data for selected compositions are listed in Table~\ref{table:two}. We find that the ideal strengths of the binary V alloys increase with increasing Cr content for all directions and decrease with Ti addition for all directions. For instance along the $[001]$ direction, an addition of $10$\% Ti to V reduces the ITS from $11.6$ GPa to $10.5$ GPa. If however $10$\% Cr are added to the V matrix, the ITS increases by $1.4\,$GPa.
Furthermore, we infer that the $[001]$ direction exhibits the lowest maximum strain value among all three directions for all V-based alloys.

The addition of chromium or titanium to the vanadium matrix leads to opposite effects in the ideal strength of binary vanadium alloys. It is apparent from our data, that the effect of both elements on $\sigma_\text{m}$ is however of similar magnitude.
As a result, $\sigma_\text{m}$ is almost unchanged if equal amounts of Cr and Ti are alloyed to V. That is, equi-composition V-Ti-Cr alloys possess nearly the same ideal strength as pure vanadium (equi-composition alloys are situated along the diagonal in Fig.~\ref{fig:alloys}). For example, the ideal strength of V-5Cr-5Ti is $11.6$ GPa, which is identical to the value of V.

Comparing the maximum strains on the primary and the secondary deformation paths along the $[001]$ direction with each other,
we can see that the orthorhombic bifurcation occurs before the maximum stresses on the primary tetragonal deformation path are reached for all selected V-based alloys. The maximum stresses of the alloys decrease due to the orthorhombic branching, e.g., the $18.4$ GPa of the binary V-10Cr alloy on the tetragonal path drops to $13.0$ GPa on the orthorhombic path.

Li \emph{et al.}~\cite{Li:2007} calculated the ideal strength of Ti-V alloys along the $[001]$ direction for concentrations of Ti $\ge 30$ at.\% based on a PP method and the VCA. Their results confirm our trend, namely that the more Ti is present in the alloy the more the ideal strength is decreased. To further assess our results, we also calculated the ideal strength on the primary deformation path of the V-30Ti alloy. Our value of $9.4$ GPa is close to the $10$ GPa obtained by Li \emph{et al.}~\cite{Li:2007} for the same alloy composition. More importantly, the alloying effect obtained by Li \emph{et al.} is $-6.7$ GPa/$30\%$Ti, which is in perfect agreement with the one obtained by us.
Assuming a linear decrease of the ITS with increasing Ti concentration in the range from 0\,\% to 30\,\%, the alloying effect on the ITS along the secondary deformation path is $-0.13$ GPa/1\%Ti as obtained by Li \emph{et al.}~\cite{Li:2007}, which is in reasonable good agreement with our value $-0.11$ GPa/1\%Ti.

From the above results, we conclude that alloying V with Ti and/or Cr produces similar trends (although quantitatively slightly different) on the ITS for two strain paths for an uniaxial load along the $[001]$ direction. A bifurcation from the primary tetragonal path to the secondary orthorhombic path accompanied by a significant reduction of the ITS occurs for all investigated V-based alloys. However, one should point out that the above results might not be the case for many other complex solid solutions, where alloying may very well induce or remove the above orthorhombic branching of the uniaxial strain and thus have more severe impact on the maximum attainable ideal stress.

\begin{table*}[tb]
\centering
\caption{\label{table:two}The ideal tensile strength $\sigma_{\text{m}}$, corresponding strain $\epsilon_{\text{m}}$ from our calculations in $[001]$, $[111]$ and $[110]$ directions for vanadium-based alloys and $[001]$ for Mo-$10$\%Tc alloy.}
\begin{ruledtabular}
\begin{tabular}{lcccccccc}
\multirow{2}{*}{composition} & \multicolumn{2}{c}{{Tetragonal}/$[001]$}& \multicolumn{2}{c}{{Orthorhombic}/$[001]$} & \multicolumn{2}{c}{direction/$[111]$} & \multicolumn{2}{c}{direction/$[110]$}\\
\cline{2-9}
    & $\sigma_{\text{m}}$ (GPa) & $\epsilon_{\text{m}} (\%)$ & $\sigma_{\text{m}}$ (GPa) & $\epsilon_{\text{m}} (\%)$ & $\sigma_{\text{m}}$ (GPa) & $\epsilon_{\text{m}} (\%)$

    & $\sigma_{\text{m}}$ (GPa) & $\epsilon_{\text{m}} (\%)$ \\
 \hline
    V-10Cr& 18.4 & 16.2&13.0&9.2 &27.7 &34.3 &38.7 &37.6 \\
    V-5Cr & 17.3 & 16.1&&& 27.3 &34.4 &38.2 &38.0\\
    V & 16.1 & 16.0&11.6&9.0 & 26.7 & 35.2 & 37.6 & 38.3 \\
    V-5Cr-5Ti & 16.0 & 16.0&11.6&9.1 &26.5  &35.3  &37.3  &38.2\\
    V-5Ti &  14.9 & 15.3 &&& 26.1& 35.2 & 36.6 & 38.5 \\
   V-10Ti &  13.7& 15.2&10.5&9.0& 25.4  &35.3 &35.9 &38.3\\
   Mo-10Tc & 24.9& 11.6 & & & &&&\\
\end{tabular}
\end{ruledtabular}
\end{table*}

\section{\label{sec:discussion}Discussion}

The anisotropy of the maximum stress of V between the $<111>$ directions and the $<001>$ directions
may be understood on the basis of structural energy differences (SEDs)~\cite{Skriver:1985,Paxton:1990,Pettifor:1995,Andersen:1985} of \emph{cubic} structures as outlined in the following. Here, we focus on the primary tetragonal deformation path (along the $<001>$ directions) because this deformation follows the Bain path from bcc to fcc. The secondary orthorhombic deformation path follows the Bain path only up to the branching point.
Stress along one of the $<111>$ directions occur along a trigonal deformation path from bcc to sc.
Using SEDs we explain the calculated alloying trend of Mo-Tc and the V-based alloys in the $<111>$ and $<001>$ directions.

We note that the presented arguments can not be used to explain the relative magnitude of $\sigma_\text{m}$ for uniaxial loading in the $<110>$ directions of the bcc lattice, because the distorted orthorhombic lattice does not coincide with a higher symmetric cubic one for any value of the strain $\epsilon>0$ (Poisson contraction), i.e., the uniaxial strain energy does not level off as if there is a nearby symmetry dictated extremum.

\subsection{\label{sec:strengthValloy} Structural energy difference and ideal strength}
The distorted bct lattice of the $[001]$ tetragonal deformation and the distorted hex lattice of the $[111]$ deformation coincide with the fcc lattice (at $c_{\text{bct}}/a_{\text{bct}}=\sqrt{2}$) and with the sc lattice (at $c_{\text{hex}}/a_{\text{hex}}=\sqrt{3/2}$), respectively. The fcc and the sc structures of V, as well as fcc Mo were identified to be the nearest symmetry-dictated maxima to the bcc phase of the respective energy versus strain curves,~\cite{Craievich:1994,Marcus:2002,paw:2,Schonecker:2011} i.e., the uniaxial strain energy must level off to the fcc-bcc SED and to the sc-bcc SED for an elongation along $[001]$ and an elongation along $[111]$, respectively, albeit at strains larger than the maximum strain ($\epsilon_{\text{m}}$).

The fcc structure of the tetragonal strain path ($[001]$ direction) and the sc structure of the trigonal strain path ($[111]$ direction) can thus be identified as the structures associated with the minimum-energy ``barrier`` in total energy~\cite{Milstein:1994,Milstein:1995} that appear on the two uniaxial loading paths of V and Mo nearby the equilibrium bcc phase.
The existence of a nearby maximum in energy on a particular strain path implies a limitation on the attainable maximum stress.~\cite{Milstein:1971,Milstein:1995,Sob:2004,Mo:1} It restricts the uniaxial strain energy ($\Delta E $) to the SEDs within the strain interval ($\Delta \epsilon$) to accomplish the transformation from bcc to either fcc or sc, i.e., the ratio $\Delta E/\Delta \epsilon$ is bounded.

It should be pointed out once more that, in the case of V or Mo, the fcc structure and the sc structure are symmetry-dictated maxima of the uniaxial strain energy along the $[001]$ primary tetragonal strain path and along the $[111]$ trigonal strain path, respectively. As a consequence, the attainable maximum stress may be directly correlated to the (structural) energy difference between these maxima and the nearby bcc equilibrium phase.
A theoretically possible alternative bcc to fcc transformation (or vice versa) via intermediate bct states may occur under the assumption that both bcc and fcc are minima in energy of the tetragonal configuration space. In this case, the minimum energy barrier for the bcc to fcc transformation on the uniaxial deformation path is associated with a stress-free bct state.~\cite{Milstein:1994,Milstein:1995} While the existence of such an unstressed intermediate state is guaranteed, no definite statements on neither the height of the minimum energy barrier nor its geometry (e.g., lattice parameter ratio, $c_{\text{bct}}/a_{\text{bct}}$) can be made without knowledge of the deformation path itself, because the lattice parameter ratio of a stress-free bct state is not dictated by symmetry.~\cite{Craievich:1994,Marcus:2002,Schonecker:2011} It appears that a simple estimation of the attainable maximum stress on the basis of SEDs is hindered in this case.

For a quantitative estimation of the ITS, we below assume a state of strain with constant volume fixed to the theoretical bcc equilibrium volume per atom, $\Omega_{\text{bcc}}$. This is to keep the presented argument both simple and general (the only additional input is the total energy of the fcc state), and is neither a severe approximation since the relaxed volumes of all three cubic structures of V, i.e., the volumes of the initial and final states of the transformation, differ from another by less than $6\,\%$.
We then may approximate the ``true'' $\sigma_\text{m}$ by rewriting Eq.~\eqref{eq:stress},
\begin{equation}
 \sigma^{\phantom{S}}_{\text{m}}\approx \sigma^{\text{SED}}_{\text{m}} = \frac{1}{\Omega_{\text{bcc}}}\frac{\Delta E}{\Delta \epsilon},
\label{eq:stress:strdiff}
\end{equation}
where $\Delta E$ is either the fcc-bcc or the sc-bcc SED, and $\Delta \epsilon$ is the strain at constant volume necessary to transform the bcc lattice into either the fcc lattice or into the sc lattice. Readily one may find, $\Delta \epsilon_{[001]} = (1-0.5^{1/3})/0.5^{1/3}\approx 0.260$ and $\Delta \epsilon_{[111]} = (1-0.25^{1/3})/0.25^{1/3}\approx 0.587$ for the $[001]$ distortion and the $[111]$ distortion, respectively.
For $\Delta E$ of V, we computed $E_\text{fcc} -E_{\text{bcc}}=20.48\text{mRy/atom}$ and $E_\text{sc}-E_{\text{bcc}}=82.76\text{mRy/atom}$. Using $\Omega_{\text{bcc}}=13.4695\,\textrm{\AA}^3$ and Eq.~\eqref{eq:stress:strdiff}, the above SEDs yield  $\sigma^{\text{SED}}_{\text{m}}=12.8\,\text{GPa}$ and $\sigma^{\text{SED}}_{\text{m}}=22.9\,\text{GPa}$  for the $[001]$ distortion and the $[111]$ distortion, respectively, their ratio being $1.79$.
These simple estimates should be compared to our ab initio results (Table~\ref{table:one}), where we obtained $16.1\,\text{GPa}$ and $26.7\,\text{GPa}$ for the respective ideal stresses, and their ratio is $1.66$.

For the V-based alloys, from Table~\ref{table:two}, we can see that the ratio of the calculated maximum strain along the $[111]$ direction to the calculated maximum strain along the $[001]$ direction for all alloys varies from 2.11 to 2.32. That seems plausible since the strain at constant volume necessary to transform the bcc lattice into the sc lattice ($\Delta \epsilon_{[111]}$) is much larger than the strain necessary to transform the bcc lattice into the fcc lattice ($\Delta \epsilon_{[001]}$), in fact their ratio is 2.260. This agreement may however be fortunate. It is obvious that the ``true'' $\epsilon_{\text{m}}$ is much smaller than $\Delta \epsilon$, since the uniaxial strain energy levels off around the stable bcc state and around the fcc and sc maxima. If this leveling off is presumably symmetric on the minimum and the maximum in energy then $\epsilon_{\text{m}}$ equals roughly $\Delta \epsilon/2$ for the respective directions, which however preserves their mutual ratio (2.260).

\subsection{\label{sec:electronicstructure}Electronic structure and ideal strength}
The ideal strength of materials is ultimately due to the atomic bonding strength. It is well established~\cite{Friedel:1969,Pettifor:1987a,Harrison:1989} that the electronic structure and the bonding in transition metals is governed by a narrow valence $d$-band that hybridizes with a broader valence nearly-free-electron $sp$-band.
SEDs for the transition metal series (disregarding the effect of magnetism in the present discussion) that give rise to the experimentally observed sequence of stable crystal structures are well understood in terms of band filling within this band picture.~\cite{Pettifor:1995,Skriver:1985,Paxton:1990}
It is clear that mainly the gradual filling of the $d$-band not only dictates crystal structure sequences in all three transition metal series, but also trends of the equilibrium volume dependence,~\cite{Harrison:1989,Moruzzi:1977} the cohesive energy,~\cite{Moruzzi:1977,Gelatt:1977,Pettifor:1995} the bulk modulus,~\cite{Moruzzi:1993} and elastic constants~\cite{Soderlind:1993,Wills:1992}.
Here we argue that the observed trend of $\sigma_{\text{m}}$ for V-Cr-Ti and Mo-Tc alloys may as well be understood on the basis of band filling arguments.

\begin{figure}[hct]
\begin{center}
\resizebox{0.9\columnwidth}{!}{\includegraphics[clip]{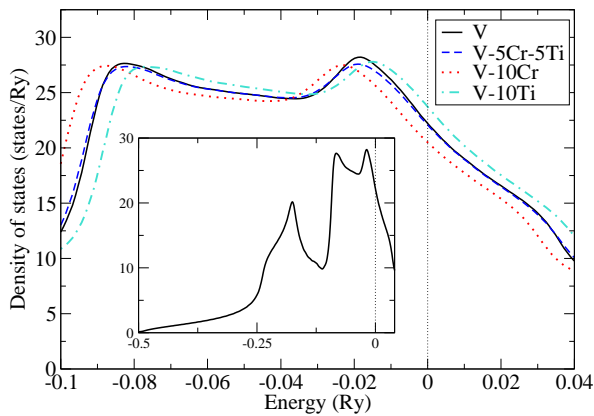}}
\caption{\label{fig:DOS}(Color online) The DOSs of V-based alloys show that alloying with a single element leads mainly to a rigid band shift of the DOS. The inset displays the complete valence band of pure V. Energies are given relative to the Fermi level which is indicated by a vertical dotted line.}
\end{center}
\end{figure}

Figure~\ref{fig:DOS} shows the density of states (DOS) of pure V, vanadium alloyed with 10\,\% Cr and with 10\,\% Ti, as well as the DOS of the equi-composition V-5Cr-5Ti alloy as obtained from our CPA calculations. We notice that the curves of V-10Cr and of V-10Ti are almost rigidly shifted with respect to the one of V, while the shape of their DOSs is merely effected by alloying vanadium with adjacent elements in the periodic table.
This rigid band shift behavior signals an increase of the $d$-occupation (increase of the number of valence electrons) and a decrease of the number of $d$-occupation (decrease of the number of valence electrons) for Cr addition and Ti addition, respectively. Thus, alloying V with Cr and Ti changes the band filling.
The DOS of V-5Cr-5Ti is virtually not shifted with respect to the one of V which indicates a zero net gain in the number of $d$-electrons in the matrix.

\begin{figure}[hct]
\begin{center}
\resizebox{0.9\columnwidth}{!}{\includegraphics[clip]{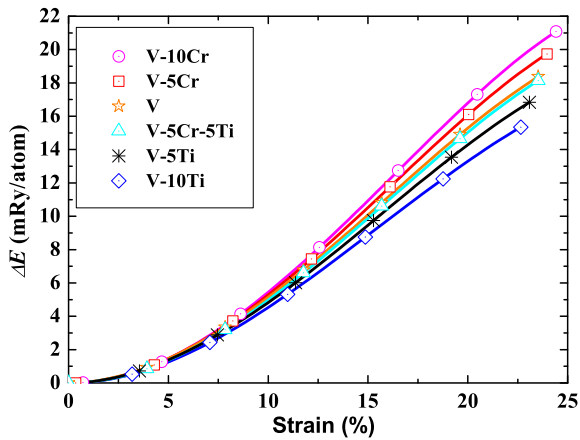}}
\caption{\label{fig:alloys energy}(Color online) The uniaxial strain energy of V-based alloys for an applied strain in the [001] direction (primary deformation path). The higher the $3d$-band occupation is the steeper the curve progression is.}
\end{center}
\end{figure}

Since the band filling is the most important parameter determining structural stability in transition metals, it is straightforward to correlate the $d$-band filling due to alloying to the ideal stress via SEDs.
For the refractory element V with a $d$-band occupation of approximately four electrons, one expects an increase of the fcc-bcc SED if more electrons are added into the $d$-band since the fcc-bcc SED of Cr is larger than the one of V.~\cite{Skriver:1985,Paxton:1990,Schonecker:2011} That is, the SEDs corresponding to intermediate band fillings are simply assumed to lie in between the values of adjacent elements, corroborated by results of canonical band theory.~\cite{Andersen:1985,Skriver:1985}
In other words, adding electrons starting at the band filling corresponding to the one of V stabilizes the bcc structure with respect to the fcc structure.
The opposite holds for Ti alloying: the fcc-bcc SED of Ti is smaller than the one of V which means a destabilization of the bcc structure with respect to the fcc structure if electrons are removed from the $d$-band. Figure~\ref{fig:alloys energy} displays the uniaxial strain energy as a function of strain for V based alloys in $[001]$ direction. It is apparent that with respect to the curve of V, the energy-strain curve rises more rapidly with increasing Cr concentration and decreases more rapidly with increasing Ti concentration. Furthermore, the energy-strain curve of the equi-composition alloy (V-5Cr-5Ti) is similar to that of V.
Based on the above trends and according to the behavior of SEDs with band filling, we expect that the ideal strength in the $[001]$ direction increases (decreases) with increasing Cr (Ti) concentration. To the best of our knowledge, sc-bcc SEDs for $3d$ transition metals have not been systematically investigated so far apart from the FM elements Fe, Co, and Ni,~\cite{Zeleny:2011} and Cu.~\cite{Wang:2003} From the trend of the sc-bcc SEDs we may however deduce the behavior of $\sigma_{\text{m}}$ along the $[111]$ direction as a function of band filling.

To confirm our expectation, we computed SEDs of vanadium based alloys and present the results in Table~\ref{table:strendiff:alloy}. Following the same reasoning as before,
we also computed $\sigma^{\text{SED}}_{\text{m}}$ according to Eq.~\eqref{eq:stress:strdiff} employing the theoretical equilibrium volumes of these bcc alloys that we reported prior~\cite{Xiaoqing}. Note that the influence of the volume change due to alloying (which enters Eq.~\eqref{eq:stress:strdiff}) is about $2\,\%$ and is not primarily responsible for the observed trend of $\sigma_{\text{m}}$ since the volume effect is much smaller than the effect of $\Delta E$.

\begin{table}[htbpc]
 \caption{\label{table:strendiff:alloy}SEDs (in mRy) of the fcc and the sc structure with respect to the stable bcc phase for bcc V$_{1-x-y}$Cr$_{x}$Ti$_{y}$ random alloys. Based on these energy differences and volumes from Ref.~\onlinecite{Xiaoqing}, the ideal strength $\sigma^{\text{SED}}_{\text{m}}$ (in GPa) is approximated according to Eq.~\eqref{eq:stress:strdiff}.}
\begin{ruledtabular}
 \begin{tabular}{lcccc}
\multirow{2}{*}{composition} & \multicolumn{2}{c}{direction/$[001]$} & \multicolumn{2}{c}{direction/$[111]$} \\
\cline{2-5}
    & $E_{\text{fcc}}-E_{\text{bcc}}$ & $\sigma^{\text{SED}}_{\text{m}}$ & $E_{\text{sc}}-E_{\text{bcc}}$ & $\sigma^{\text{SED}}_{\text{m}}$ \\
\hline
V-10Cr & 23.00 & 14.6 & 84.79 &  23.9 \\
V-5Cr  & 21.74 & 13.7 & 83.99 & 23.5 \\
V & 20.48 & 12.8 & 82.76 & 22.9 \\
V-5Cr-5Ti & 20.28 & 12.6 & 82.76 & 22.8 \\
V-5Ti & 19.06 & 11.8 & 81.98 & 22.4 \\
V-10Ti & 17.63 & 10.8 & 80.81 & 21.9
\end{tabular}
\end{ruledtabular}
\end{table}

Our calculated SEDs clearly substantiate an increase of $E_\text{fcc}-E_{\text{bcc}}$ with Cr addition, while the opposite effect is observed upon alloying with Ti. The function $E_\text{sc}-E_\text{bcc}$ follows the same trend. On the basis of our simple estimate of the ideal strength (Table~\ref{table:strendiff:alloy}), we conclude with respect to bulk V an increase of $\sigma_{\text{m}}$ for the V-Cr binary alloys and a reduction of the ideal strength in case of V-Ti binary alloys. Our data affirms the larger relative change of $\sigma_{\text{m}}$ for the $[001]$ direction compared with the relative change of $\sigma_{\text{m}}$ for the $[111]$ direction.
We find for the equi-composition alloy V-5Cr-5Ti, that its ideal strength almost retains the ideal strength of pure V.

Here we would like to bring up an interesting and important detail related to the above correlations. For the $[001]$ distortions in V-Cr-Ti alloys, it is found that the composition dependence of the ideal strength correlates well with that of the SEDs and also with the one followed by $E_{<001>}$, which is ultimately determined by the composition dependence of $C'$.~\cite{Xiaoqing} However, for the $[111]$ distortions, the ideal strength vs.\! SED correlation is not followed by the $E_{<111>}$. For instance, Cr increases the $[111]$ ideal strength and the sc-bcc energy difference, but slightly decreases $E_{<111>}$ and $C_{44}$. In other words, the present results confirm the often quoted $C'$ vs.\! fcc-bcc energy difference correlation.~\cite{Wills:1992,Soderlind:1993} However, no such correlation seems to be valid between the $C_{44}$ (or $E_{<111>}$) and the sc-bcc energy difference.

For Mo-Tc alloys, Tc decreases the ideal strength compared to the value of Mo. The reason is that bcc Mo has with approximately $4.6$ electrons, as obtained from our EMTO calculation, a larger $d$-band occupation than V. According to band filling theory and DFT calculations,~\cite{Skriver:1985,Schonecker:2011} the fcc-bcc SED of Tc is smaller than the one of Mo which means a destabilization of the bcc structure with respect to the fcc structure. Thus, based on Eq.~\eqref{eq:stress:strdiff}, the ideal strength of Mo-Tc should be smaller than that of Mo, which is in line with the results from Fig.~\ref{fig:Mo}.

\section{Conclusions}

Using the EMTO method in combination with the CPA, we have investigated the ideal strength of bcc V-based alloys in the $[001]$, $[110]$, and $[111]$ directions and of Mo in the $[001]$ direction. In the $[001]$ direction, we calculated the ITS of V and V-based alloys along the tetragonal and orthorhombic deformation paths. All other ideal strengths considered in this work correspond to isotropic Poisson contraction, viz. no shape change was allowed perpendicular to the uniaxial stress.
The present computed values for both V and Mo are in good agreement with previous calculations and thus confirm that our methodology has the accuracy needed for such kind of calculations.

For Mo-Tc random alloy, we have found that adding Tc to Mo decreases its ITS.
 In the case of V-based alloys, we have obtained that Cr addition increases and Ti addition decreases the ideal strength of bcc V in all three crystallographic directions.
As a consequence, the ideal strength of V-Cr-Ti alloys remains virtually unchanged if equal amounts of Cr and Ti are introduced into the V host. We have shown that the observed alloying effects on the ideal strength can be understood on the basis of band filling arguments and structural energy differences.

The strength of a material in the range of validity of Hooke's law is characterized by its elastic constants. In our previous paper~\cite{Xiaoqing}, we investigated the strengthening effects of Cr and Ti on the elastic constants of disordered V-Cr-Ti alloys and found that Cr increases the tetragonal shear modulus ($C'$) and Ti decreases it, while alloys with equi-concentrations have similar $C'$ values as the elemental V. The effects of Cr and Ti on the $C_{44}$ elastic constant of V are somewhat smaller (and have opposite signs) compare to those obtained for $C'$. On this ground, combining our findings for the elastic regime and those obtained in this work for the ideal strength of V-Cr-Ti ternary alloys, we conclude that Cr strengthens the ITS and $C'$ of these V-based alloys and Ti has a weakening effect on both ITS and $C'$.
Furthermore, it is known,~\cite{Wills:1992,Soderlind:1993} that the $E_\text{fcc}-E_{\text{bcc}}$ SEDs in nonmagnetic transition metals scale with their $C'$. This relation can now be extended to our vanadium alloys: an increase of the $d$-band filling by alloying Cr to V leads to increasing $C'$ and ideal stress along the $[001]$ direction, which scales with the corresponding SED. The effect of adding Ti to the V matrix (decrease of the $d$-band filling) is opposite to that of Cr addition on both $C'$ and $\sigma_{\text{m}}$. We have shown that no similar correlation between $C_{44}$, ideal strength and sc-bcc structural energy difference can be established for the $[111]$ mode.

The present results offer a consistent starting point for further theoretical modeling of the micro-mechanical properties of technologically important transition metal alloys. Based on these achievements, we conclude that the EMTO-CPA approach provides an efficient and accurate theoretical tool to design the mechanical strength of nonmagnetic bcc random solid solutions and reveal the composition dependence of this fundamental physical parameter. Nevertheless, in such applications one should always monitor the basic muffin-tin and single-site CPA errors and make sure that they remain at acceptable level as a function of the lattice distortion and chemical composition.  The extension of the present investigation to the case of magnetic alloys (such as the stainless steels) and to other Bravais lattices is in progress.

\section{ACKNOWLEDGEMENTS}
The Swedish Research Council, the Swedish Steel Producers' Association, the European Research Council, the China Scholarship Council and the Hungarian Scientific Research Fund (research project OTKA 84078), and the National Magnetic Confinement Fusion Program of China (2011GB108007) are acknowledged for financial support. S.\! S. gratefully acknowledges the Carl Tryggers Stiftelse f\"or Vetenskaplig Forskning and Olle Erikssons stiftelse f\"or materialteknik.


%

\end{document}